\documentclass{elsart}
\usepackage{graphics}
\def\eref#1{(\protect\ref{#1})}
\def\fref#1{\protect\ref{#1}}
\def\sref#1{\protect\ref{#1}}
\def\tref#1{\protect\ref{#1}}
\def\etal{{\it{}et~al.}}

\newdimen\captwidth
\newdimen\normalfigwidth
\normalfigwidth=\captwidth                     % width figures appear

\def\normalfigure#1{\hbox to\textwidth{%
       \hfil\resizebox{\normalfigwidth}{!}{\includegraphics{#1}}\hfil}}

\def\capt#1{\refstepcounter{figure}\bigskip\hbox to \textwidth{%
       \hfil\vbox{\hsize=\captwidth\renewcommand{\baselinestretch}{1}\small
       {\sc Figure \thefigure}\quad#1}\hfil}\bigskip}

\def\tcapt#1{\refstepcounter{table}\bigskip\hbox to \textwidth{%
       \hfil\vbox{\hsize=\captwidth\renewcommand{\baselinestretch}{1}\small
       {\sc Table \thetable}\quad#1}\hfil}\bigskip}

\begin{document}

\date{}
\journal{the Proceedings of the Royal Society, Series B}

\begin{frontmatter}
\title{Extinction, diversity and survivorship of taxa in the fossil record}
\author{M. E. J. Newman}
\address{Santa Fe Institute, 1399 Hyde Park Road, Santa Fe, NM 87501.
U.S.A.}
\author{Paolo Sibani}
\address{Fysisk Institut, Odense Universitet, DK 5230 Odense M.   Denmark}
\vspace{1in}
\begin{abstract}
  Using data drawn from large-scale databases, a number of interesting
  trends in the fossil record have been observed in recent years.  These
  include the average decline in extinction rates throughout the
  Phanerozoic, the average increase in standing diversity, correlations
  between rates of origination and extinction, and simple laws governing
  the form of survivorship curves and the distribution of the lifetimes of
  taxa.  In this paper we derive mathematically a number of relations
  between these quantities and show how these different trends are
  inter-related.  We also derive a variety of constraints on the possible
  forms of these trends, such as limits on the rate at which extinction may
  decline and limits on the allowed difference between extinction and
  origination rates at any given time.
\end{abstract}
\end{frontmatter}

\newpage

\section{Introduction}
\label{intro}
Traditionally, studies of origination and extinction have focussed on
single taxa or groups of taxa, in an attempt to determine the causes of
particular extinction events.  In the latter half of the twentieth century,
the larger mass extinction events have also drawn a considerable amount of
attention from researchers curious about their scope and origin.  However,
it has only been in the last twenty-five years or so that fossil data have
been available of sufficient quality to allow statistical investigation of
large-scale patterns in the extinction and origination record.  One of the
earliest statistical studies was that of Van Valen~(1973), who looked at
``survivorship curves'' for a large number of different groups of organisms
and conjectured possible mechanisms to explain what he saw.  More recently,
a variety of authors have looked at other trends in the fossil record, and
proposed simple empirical laws governing various aspects of taxon
origination and extinction.  Of particular interest to us in this paper, in
addition to Van Valen's work, will be work on the distribution of the
lifetimes of taxa (Sneppen~\etal~1995), the increase in diversity towards
the recent (Benton~1995), and the decline in extinction rates
(Sibani~\etal~1995, Newman and Eble~1998).

Our concern in this paper is with overall statistical trends in the fossil
record of the last 600 million years~(My) or so.  For instance, it has been
observed that the mean extinction rate for genera (or families) appears to
decline during the Phanerozoic.  There is certainly a great deal of
fluctuation about this mean, with many intervals of time in which
extinction increases rather than falls off.  We however will look only at
the general trend and not at these fluctuations.  Thus our work does not
for example address the appearance of mass extinction events in the record,
or such controversial issues as the causes of particular events or the
number of taxa killed.

In this paper we present a mathematical treatment of the trends in the
fossil record and show that they are not independent but are in fact
closely related to one another.  Starting with only a few very basic and
uncontentious observations about patterns of extinction and diversity, we
derive quite strict limits on the way in the fossil record can behave.  We
show for instance that the origination rate for families must be equal to
the extinction rate at all times to within about one family per million
years---a similar result pertains for genera or any other taxonomic
subdivision---and that although the average extinction rate declines during
the Phanerozoic, there is a limit on how fast it can decline.  These and
other conclusions are derived in Section~\sref{examples}.

\begin{table}[t]
\begin{center}
\begin{tabular}{c|c|c|c}
\hline
\hline
variable & meaning & conjectured form & reference \\ 
\hline
$x(t)$      & extinction rate       & $t^{-1}$                & Newman \& Eble~1998 \\
$y(t)$      & origination rate      & declines                & Van Valen \& Maiorana~1985 \\
$p(t)$      & percent extinction    & $t^{-1}$                & Sibani~\etal~1995 \\
$D(t)$      & standing diversity    & $\exp(t/t_D)$           & Benton~1995 \\
$w_t(\tau)$ & survivorship          & $\exp[-\tau/\tau_w(t)]$ & Van Valen~1973 \\
$R(\tau)$   & lifetime distribution & $\tau^{-\beta}$         & Sneppen~\etal~1995 \\
\hline
\hline
\end{tabular}
\end{center}
\begin{center}
\tcapt{The various quantities appearing in our theory.}
\end{center}
\label{vars}
\end{table}

\begin{figure}
\normalfigure{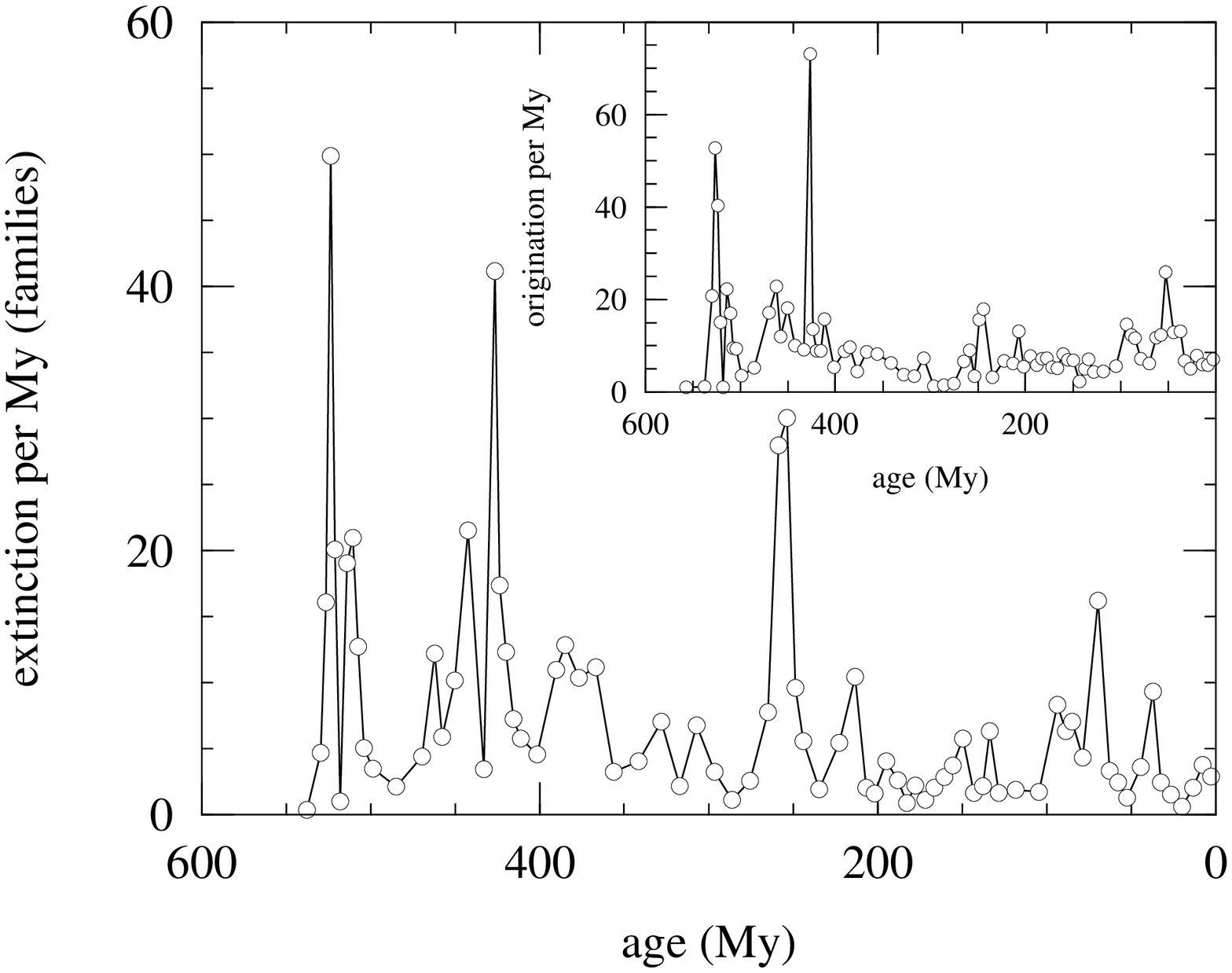}
%\centerline{\psfig{figure=decline.ps,height=10cm,width=12cm}}
\capt{Main figure: extinction rate in families per million years for marine
  organisms as a function of time.  Inset: origination rate for marine
  organisms.  Data are from the compilation by Sepkoski~(1992).}
\label{decline}
\end{figure}

\begin{figure}
\normalfigure{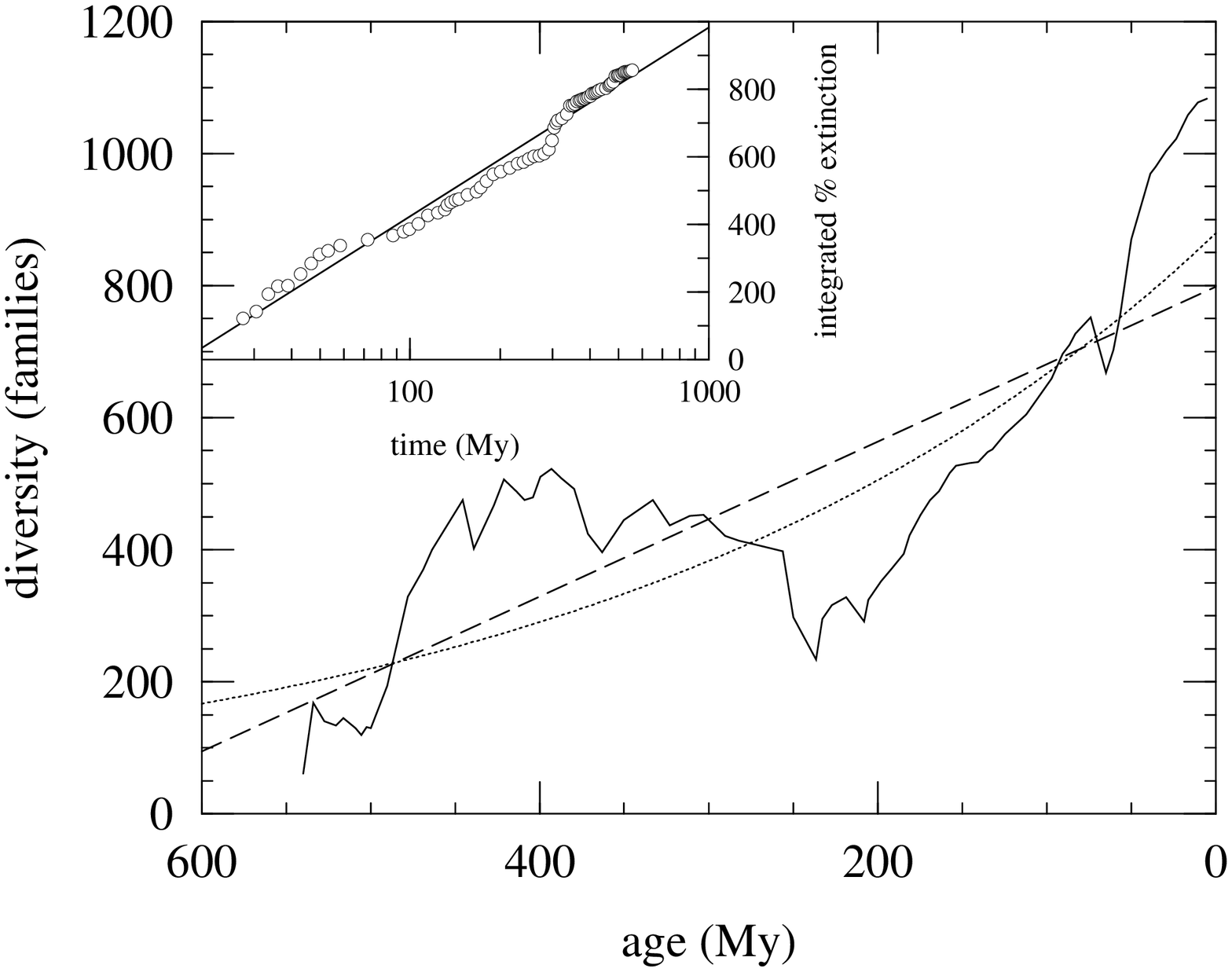}
%\centerline{\psfig{figure=diversity.ps,height=10cm,width=12cm}}
\capt{Main figure: the number of known families of organisms as a function
  of time during the Phanerozoic.  The dotted (curved) line is the best fit
  exponential and the dashed (straight) one is the best linear fit.  Inset:
  the integrated percentage extinction of families.  Data from
  Sepkoski~(1992).}
\label{diversity}
\end{figure}

\begin{figure}
\normalfigure{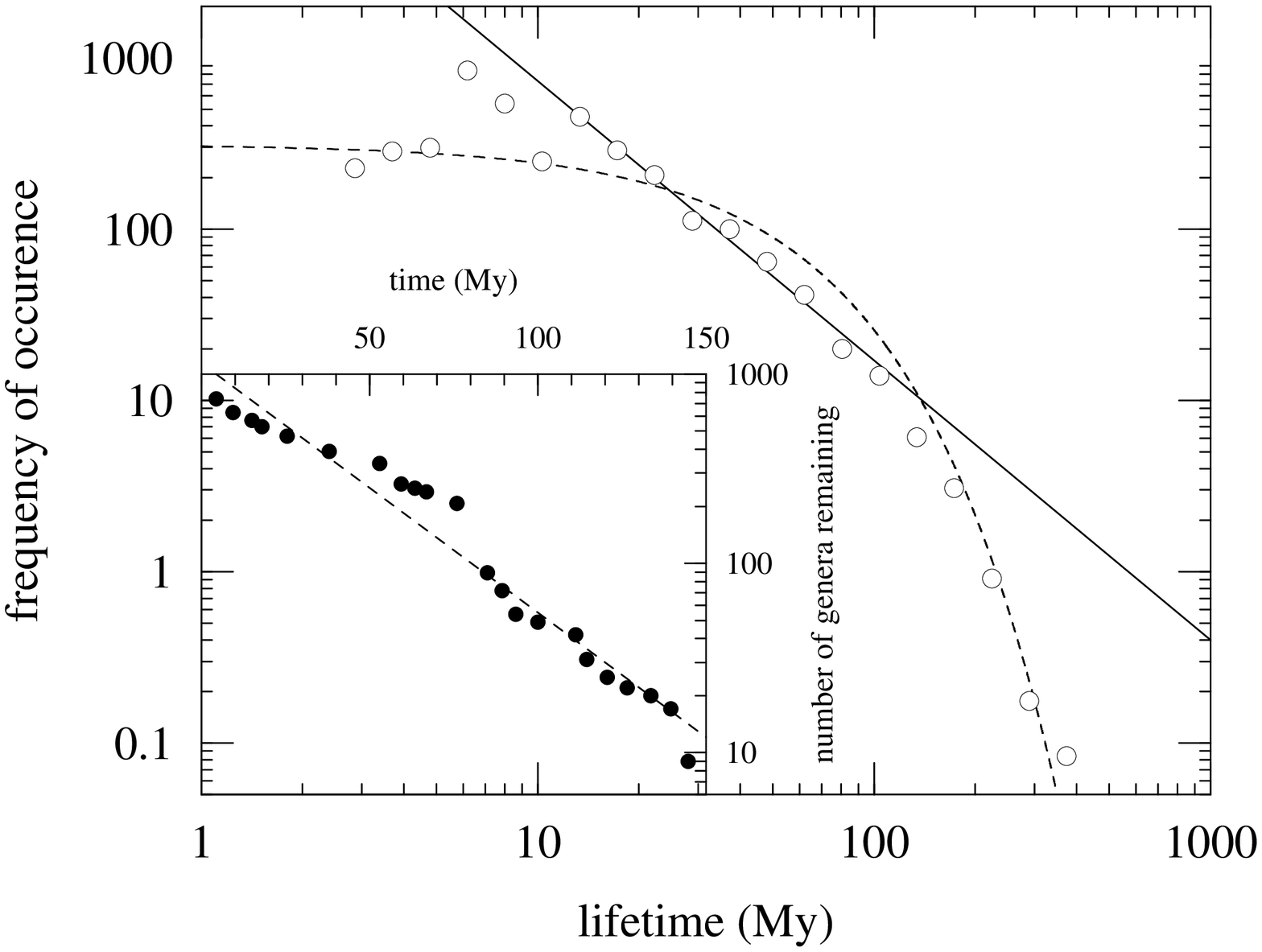}
%\centerline{\psfig{figure=lifetimes.ps,height=10cm,width=12cm}}
\capt{Main figure: Histogram of the lifetimes of marine genera during the
  Phanerozoic.  The solid line is the best power-law fit to the points
  between 10 and 100~My.  The dotted line is the best exponential fit to
  all the points.  Inset: the number of genera of marine organisms
  surviving out of an initial group of 740, over a period of 150~My up to
  the present.  The dotted line is the best fit exponential.  Data from
  Sepkoski~(1992).}
\label{lifetimes}
\end{figure}

Table~\tref{vars} lists the quantities which will appear in our discussion
of extinction and origination, along with some hypotheses about their
behaviour based on fossil evidence.
\begin{enumerate}
\item We denote by $x(t)$ and $y(t)$ the number of taxa becoming extinct
  and originating per unit time, respectively, as a function of time $t$.
  As mentioned above, it is widely believed that the extinction rate has
  declined on average during the Phanerozoic.  This decline is clear in
  Figure~\fref{decline}, which shows the extinction rate for marine
  families, drawn from an updated version of the compilation of fossil data
  by Sepkoski~(1992).  It has been suggested that on average this decline
  takes the form $x(t) \sim 1/t$, provided the origin of time is suitably
  chosen (Newman and Eble~1998).  For origination the case is less clear,
  but again it appears (Van Valen and Maiorana~1985, Sepkoski~1998) that
  the rate is falling off over time (see the inset to
  Figure~\fref{decline}).
\item Frequently one measures the fractional (or percentage) extinction
  rate in the fossil record, which is the ratio of the number of taxa
  becoming extinct per unit time to the total number in existence.  Another
  way of looking at this quantity is as the probability per unit time of
  the extinction of any one taxon.  We denote this quantity $p(t)$.  By
  studying its time-integrated value, Sibani~\etal~(1995) determined that
  $p(t)$ declines approximately according to a power law $p(t) \sim
  t^{-\alpha}$.  The most recent data indicate that the integrated form
  grows roughly logarithmically in time (see the inset to
  Figure~\fref{diversity}), and hence that $\alpha=1$.  The data are also
  compatible with slightly smaller values of $\alpha$ in the region of
  $0.8$ or $0.9$.
\item The variable $D(t)$ represents the number of taxa alive at time $t$,
  the so-called ``standing diversity''.  Overall, it appears from the
  fossil record that this quantity has increased towards the recent.
  Benton~(1995) has suggested that it does so exponentially $D(t) \sim
  \exp(t/t_D)$, although there is sufficient variation in the data that
  they can also be reasonably well fit by a linear form $D(t) \sim a + bt$.
  The diversity is shown in Figure~\fref{diversity} using data from the
  compilation by Sepkoski~(1992) again.
\item The variable $w_t(\tau)$ represents the ``survivorship curve''---the
  probability that a taxon which was alive at time $t$ is still alive a
  time $\tau$ later.  Van Valen~(1973) suggested on the basis of fossil
  data analyses that this curve takes the form $w_t(\tau) \sim
  \exp[-\tau/\tau_w(t)]$ for genera or families within single orders or
  phyla.  In other words, it falls off exponentially in $\tau$ with a rate
  constant $\tau_w$ which depends on $t$.  ($\tau_w(t)$ is also equal to
  the mean taxon lifetime at time $t$.)  This result is often referred to
  as Van Valen's law.  As Raup~(1991) has shown, this result appears to be
  true also for larger groups.  In the inset to Figure~\fref{lifetimes} we
  show the survivorship curve for all known genera of marine organisms
  which were alive 150 million years ago~(Ma).  As we can see it follows
  the predicted exponential form quite closely.  Van Valen's law has gained
  theoretical support from a model of extinction advanced by Raup~(1978)
  and by Baumiller~(1993).  This model is based on a simple
  time-homogeneous birth/death process for the species within a genus and
  contains two parameters $\lambda$ and $\mu$, which measure the rates at
  which species appear and disappear.  In the case where $\lambda\ll\mu$,
  the survivorship curve takes the form postulated by Van Valen.  On the
  other hand if $\lambda\simeq\mu$ then the survivorship curve takes the
  form $w_t(\tau) \sim 1/(1+\lambda \tau)$.
\item The variable $R(\tau)$ denotes the probability distribution of
  species lifetimes $\tau$ averaged over the entire fossil record.  It has
  been suggested (Sneppen~\etal~1995) that this distribution has a
  power-law form $R(\tau) \sim \tau^{-\beta}$, with an exponent $\beta$
  somewhere in the range from $\frac32$ to $2$, although the data are
  probably just as well fit by an exponential (see
  Figure~\fref{lifetimes}).
\end{enumerate}

We should emphasize that we are not making any statement here about the
causes of the trends described above.  The mathematical developments of the
following sections are equally valid whether the trends are the result of
biological effects such as true extinction or pseudoextinction or are
artifacts of things like taxon sorting, preservational bias or taxonomic
classification (Raup and Sepkoski~1982, Flessa and Jablonski~1985,
Pease~1992, Gilinsky~1994).  Our theory makes statements only about the
appearance and disappearance of taxa in the fossil record, and not about
causes.

In this paper we present an analysis of the relationships between the
quantities in Table~\tref{vars}, and discuss the implications of the
conjectures listed in the third column.  Our theory makes only two quite
simple assumptions.
\begin{enumerate}
\item We assume that on average the extinction rate is not increasing and
  that standing diversity is not decreasing.  These results seem to be
  widely accepted in the palaeontological community and are well supported
  by fossil data.
\item We assume that all species ultimately become extinct.  Empirically,
  this seems to be a reasonable assumption.  The mean lifetime of a species
  in the fossil record is in the range from 4 to 11~My (Valentine~1970,
  Raup~1991), and there are no known examples of species which have lived
  longer than 100~My.
\end{enumerate}

The outline of this paper is as follows.  In Section~\sref{theory} we
derive a number of relations between the quantities defined in
Table~\tref{vars} and specify under what conditions these relations should
hold.  In Section~\sref{examples} we apply our theory to the empirical
trends noted above, and explain the implications of our analysis for fossil
extinction and origination.  In Section~\sref{concs} we give our
conclusions.

\section{Theoretical framework}
\label{theory}
Consider then a record of the origination and extinction of a large number
of taxa spanning a time interval from an earliest time $t_0$ to a latest
time $t_1$.  If our time interval is the entire Phanerozoic, for instance,
then $t_0 \simeq 545$~Ma and $t_1$ is the present.  We will write our
theory in terms of general ``taxa''.  Exactly the same relations between
quantities apply whether these taxa are species, or some proxy for species,
such as genera or families, or still higher taxonomic groups such as
classes or orders.

\subsection{Standing diversity and extinction probability}
\label{divsec}
Given a time-varying taxon extinction rate $x(t)$ and origination rate
$y(t)$, as defined in Section~\sref{intro}, the standing diversity of taxa
$D(t)$ is simply the difference between the integrated number of
originations since the start of our time interval, and the integrated
number of extinctions.  In other words
\begin{equation}
D(t) = \int_{t_0}^t [y(t')-x(t')]\d t'.
\label{sd}
\end{equation}
And the fractional (or percentage) extinction per unit time $p(t)$ is the
ratio of the rate of extinction to the number of extant species:
\begin{equation}
p(t) = {x(t)\over D(t)}.
\label{killrate}
\end{equation}
As we mentioned, $p(t)$ may also be regarded as the probability per unit
time of the extinction of any one species.  Assumption~(1) from
Section~\sref{intro} implies that $p(t)$ must be a constant or decreasing
function of time.

\subsection{Survivorship curve}
\label{surv}
The survivorship curve $w_t(\tau)$ is related to $p(t)$ by the differential
equation
\begin{equation}
{\d w_t\over\d\tau} = -p(t+\tau) w_t(\tau),
\label{wdiff}
\end{equation}  
with the initial condition $w_t(0)=1$ for all values of $t$.  Essentially,
this equation says that out of a given initial group of taxa, the number
becoming extinct in any small interval of time is equal to the number still
alive multiplied by the probability of any one of them becoming extinct in
that interval.  The solution of this equation is
\begin{equation}
w_t(\tau) = \exp\Biggl[-\int_t^{t+\tau} p(t')\d t'\Biggr].
\label{survivorship1}
\end{equation}
Now our second assumption from Section~\sref{intro} comes into play.  In
order that all species ultimately die, we require that $w_t(\tau)\to0$ as
$\tau\to\infty$, which is only possible if the integral in
Equation~\eref{survivorship1} diverges in the same limit.  The
fastest-decaying functional form for $p(t)$ which gives a divergent
integral is\footnote{In fact, we can include a succession of factors of the
  form $\log\log t$, $\log\log\log t$ and so on in the denominator of the
  function and still get a divergent integral.  However, these are very
  slowly varying terms and, since the available fossil data span less than
  two decades in $t$, it is safe to ignore such possibilities.}
$p(t)\sim1/(t\log t)$.  We can however eliminate the factor of $\log t$
from this expression by appealing to only the grossest features of the
fossil record, because if $p(t) \sim 1/(t \log t)$, then
Equation~\eref{survivorship1} implies that
\begin{equation}
w_t(\tau) = {\log t\over\log(t+\tau)}.
\end{equation}
This is a very slowly decaying function of $\tau$, and not remotely similar
to the approximately exponential survivorship curves seen in the fossil
record.  It is therefore fairly safe to say that $p(t)$ should decay no
faster than $1/t$.  Coupled with the results of Section~\sref{divsec} this
means that $p(t)$ must lie somewhere between being constant in time and
falling off as $1/t$.  This in turn implies that the extinction rate $x(t)$
also lies between the same limits, and that the diversity $D(t)$ lies
somewhere between being constant in time and growing as $t$.  These are
fairly stringent constraints on these quantities, and allow us to rule out
many possible functional forms.

The origin of time $t=0$ in these expressions is not fixed by the analysis,
and may be varied in any way we desire to fit the fossil data, with the
qualification that it must fall before the earliest time $t_0$ for which
data are available.
%Newman and Eble~(1998), for example, found that
%extinction data take a particularly simple form when the origin of time is
%set about $260$~My before the start of the Cambrian.

Our limits on the form of $p(t)$ also place limits on the survivorship
curve.  The fastest decreasing form of $w_t(\tau)$ corresponds to the case
of $p(t)$ constant in time.  In that case, we find that
\begin{equation}
w_t(\tau) = \exp\biggl[-{\tau\over\tau_w}\biggr],
\label{survivorship2}
\end{equation}
with a constant mean taxon lifetime $\tau_w=1/p(0)$, where $p(0)$ is the
constant value of the extinction probability.  This form however seems
unlikely as far as the Phanerozoic fossil record is concerned, since a
constant value of $p(t)$, although compatible with our theory, would imply
that both the extinction rate $x(t)$ and the diversity $D(t)$ are also
constant (see Equation~\eref{sd}), which is not in very good agreement with
the data.  In fact, as Figure~\fref{diversity} shows, the best available
data indicate that $p(t)$ is quite close to its lower limiting form
\begin{equation}
p(t) = {C\over t},
\label{oneovert}
\end{equation}
for some constant $C$.  This form corresponds to the slowest decaying form
for the survivorship curve:
\begin{equation}
w_t(\tau) = \biggl({t\over t+\tau}\biggr)^C.
\label{survivorship3}
\end{equation}
Interestingly, for $C=1$, this is similar to the form postulated by
Baumiller~(1993) which we mentioned in Section~\sref{intro}, although it
arises here through a completely different mechanism.

\subsection{Origination rate}
\label{origination}
In order that the diversity $D(t)$ be a constant or increasing function of
time, we require that the integrand in Equation~\eref{sd} be non-negative.
This in turn implies that the origination rate $y(t)$ must always be
greater than or equal to the extinction rate $x(t)$.  On the other hand,
since, as we showed in the last section, the diversity can increase no
faster than linearly in time $t$, the integrand must also be a constant or
decreasing function of time.  This implies that $y(t)$ must always be less
than $x(t)$ plus some constant.  In other words, the origination rate is
constrained to lie in the range
\begin{equation}
x(t) \le y(t) \le x(t) + \epsilon,
\label{origlim}
\end{equation}
where the constant $\epsilon$ is given by
\begin{equation}
\epsilon = y(t_0) - x(t_0).
\label{defseps}
\end{equation}
If $\epsilon$ is very large compared with the typical values of $x(t)$ over
our data set then this is not a very severe constraint on the origination
rate, and $y(t)$ can still vary in almost any way it pleases.  However, as
we will show in Section~\sref{examples} when we come to examine the fossil
data in more detail, this is not the case.  In fact the value of $\epsilon$
is quite small by comparison with $x(t)$ which means that the origination
is obliged to follow the extinction rate closely.

\subsection{Lifetime distribution}
Finally, let $r_t(\tau)$ be the probability density that a species present
at time $t$ lives on for an additional time $\tau$ and then dies.  We have
\begin{equation}
r_t(\tau) = -{\d w_t\over\d\tau} = w_t(\tau) p(t+\tau).
\label{condlife}
\end{equation}
The lifetime distribution, averaged over our entire time interval, is then
given by
\begin{equation}
R(\tau) = {\int_{t_0}^{t_1-\tau} r_{t'}(\tau) y(t') \d t'\over
           \int_{t_0}^{t_1-\tau} y(t') \d t'}.
\label{lifetime}
\end{equation}
Since one usually has $\tau\ll t_1$, the $\tau$-dependence of the upper
limit of the integrals can be safely ignored.

\section{Implications for fossil data}
\label{examples}
In Section~\sref{intro} we discussed a number of different conjectures
which have been put forward about extinction and origination in the fossil
record.  We now consider the implications of each of these conjectures
within the theoretical framework presented above.

\subsection{Form of the decrease in the extinction probability}
Sibani~\etal~(1995) have conjectured on the basis of fossil data that the
extinction probability $p(t)$ per unit time decreases according to a power
law $p(t) \sim t^{-\alpha}$ with $0\le\alpha\le1$.  As discussed in
Section~\sref{intro}, our best estimate of the exponent $\alpha$ is around
1 or a little less.  This result is in good agreement with our theory.

\subsection{Form of the decrease in extinction rate}
\label{decrease}
Newman and Eble~(1998) have conjectured that the extinction rate $x(t)$ is
also falling off with time as $x(t) \sim 1/t$.  This result is compatible
with Equations~\eref{killrate} and~\eref{oneovert}, although it represents
the extreme case in which diversity is not increasing.  It may be that the
actual extinction rate decreases more slowly than this, allowing diversity
to increase as well.

\subsection{Form of the origination rate}
\label{origin}
The inequality~\eref{origlim} places limits on the values that the
origination rate can take, forcing it to follow the values of the
extinction rate, within a margin set by the constant $\epsilon$.  The value
of this constant is given by Equation~\eref{defseps}, but is rather hard to
calculate since the average extinction and origination rates $x(t_0)$ and
$y(t_0)$ at the beginning of the data set are poorly defined.  A better way
to estimate $\epsilon$ is to note that it is also equal to the slope of the
diversity curve as a function of time (Figure~\fref{diversity}) at the
point $t=t_0$.  In the case of the marine families in the Sepkoski
database, for example, this gives a value of $\epsilon = 1.4\pm0.1$
families per~My.  The average extinction rate for the same data set is
$6.1$ families per~My.  Thus, the origination rate must follow the
extinction rate quite closely at all times and can on average exceed the
extinction rate by at most about 25\%.  This in turn implies that if the
extinction rate is falling off over time (as it appears to be) then the
origination rate must, at least on average, fall off as well, and by a
similar amount and in a similar fashion.  The results are even more
pronounced in the case of genera, where $\epsilon=3.7\pm0.4$ genera per~My
while the mean extinction rate is $52.3$ genera per~My.

\subsection{Form of the increase in diversity}
Benton~(1995) has suggested that the standing diversity $D(t)$ increases
exponentially with time.  This is clearly at odds with our result that
$D(t)$ can at most increase linearly in time.  However, as
Figure~\fref{diversity} shows, the data on diversity are sufficiently poor
as to be compatible with a linear fit to within the available accuracy.

\subsection{Form of the survivorship curve}
Van Valen~(1973) suggested that survivorship curves follow an exponential
decay law $w_t(\tau) \sim \exp[-\tau/\tau_w(t)]$ within individual orders
or phyla and, as we showed in Section~\sref{intro}, a similar form seems to
hold for survivorship curves for larger groups of taxa as well.  The
results of Section~\sref{surv} show that the survivorship curve can take
precisely this form, with a constant mean taxon lifetime $t_w$, if the
percentage extinction rate $p(t)$ is a constant in time.  However, as we
pointed out, the fossil data indicate that $p(t)$ is not constant in time,
but rather is close to its lower limiting form of $p(t) = C/t$
(Equation~\eref{oneovert}).  This gives the form~\eref{survivorship3} for
the survivorship curve, rather than the exponential form postulated by Van
Valen.  This discrepancy is less serious than it may appear to be, however.
Van Valen postulated his law on the basis of a large number of plots of
survivorship curves similar to Figure~\fref{lifetimes}, which appear
straight on semi-logarithmic scales.  If we take the logarithm of
Equation~\eref{survivorship3}, we find that
\begin{equation}
\log w_t(\tau) = -C\log(1+\tau/t) \simeq -C {\tau\over t},
\label{snlaw}
\end{equation}
when $\tau\ll t$.  If the typical lifetimes $\tau$ of species are
significantly less than $t_0$, which will always be the case provided the
origin of time is well before the beginning of our data set $t_0$, then
this inequality is always satisfied and we will indeed see survivorship
curves which appear as straight lines on a semi-logarithmic plot.  Thus,
Van Valen's law appears to be compatible with our theory.

Equation~\eref{snlaw} implies that the mean lifetime $\tau_w(t)$ which
appears in Van Valen's law should increase according to
\begin{equation}
\tau_w(t) = {t\over C}.
\end{equation}
Although this particular form has not to our knowledge been conjectured
previously, there is evidence that lifetimes increase, on average, during
the Phanerozoic.  Raup~(1988) for instance has analysed the survivorship
curves of groups of fossil marine genera drawn from the Sepkoski database
and shown that mean lifetimes approximately double from the Cambrian to the
Recent.

\subsection{Form of the lifetime distribution}
As discussed in Section~\sref{intro}, the lifetimes of taxa in the fossil
record have a highly left-skewed distribution which, as
Sneppen~\etal~(1995) have suggested, may be a power law with an exponent
$\beta$ in the range from $\frac32$ to $2$.  The data are also compatible
with forms which decrease faster than a power law.  We now show that both
of these hypotheses are compatible with the current theoretical framework,
depending on whether the typical values of $t$ are larger or smaller than
the typical taxon lifetimes.

The overall distribution of taxon lifetimes is given by
Equation~\eref{lifetime}.  If we assume the form~\eref{oneovert} for
$p(t)$, which seems to be implied by our theory and the fossil data, then
$w_t(\tau)$ takes the form of Equation~\eref{survivorship3} and $R(\tau)$
is
\begin{equation}
R(\tau) \sim \int_{t_0}^{t_1} y(t) \biggl({t\over t+\tau}\biggr)^C
             {1\over t+\tau} \d t,
\label{life}
\end{equation}
where we have neglected the normalizing factor in the denominator of
Equation~\eref{lifetime} and the $\tau$-dependence of the upper limit of
the integral.  If $y(t)$ decays sufficiently fast and $\tau\gg t$ in the
region where $y(t)$ is significantly greater than zero, then this
expression becomes
\begin{equation}
R(\tau) \sim \tau^{-(C+1)} \int_{t_0}^{t_1} y(t) t^{C+1} \d t.
\end{equation}
The constant $C$ must be greater than zero, but can otherwise take any
value we please, and hence $R(\tau)$ varies as $\tau^{-\beta}$ with an
exponent $\beta>1$.  The assumption $\tau\gg t$ is not a realistic one
however.  In Section~\sref{decrease} we made the opposite and much more
plausible assumption that $\tau\ll t$ for all values of $t$.  In this case
Equation~\eref{life} becomes
\begin{equation}
R(\tau) \sim \int_{t_0}^{t_1} y(t) \biggl[1-(C+1){\tau\over t}\biggr] \d t
           = A - B\tau,
\end{equation}
where $A$ and $B$ are positive constants whose exact value depends on the
form of $y(t)$.  On the logarithmic scales of Figure~\fref{lifetimes}, this
distribution is a highly convex function---the fossil lifetime distribution
is less convex than this.  For the case where $\tau$ is neither very much
less than nor very much greater than $t$, the distribution of lifetimes can
be expected to lie somewhere in between the two extremes derived here.  We
have confirmed this expectation by numerical integration of
Equation~\eref{life} for a variety of different choices of $y(t)$.  Our
best guess is that the fossil data lie somewhere in this intermediate
regime.  This in turn implies that the correct origin of time for our
theory is earlier than the beginning of the Phanerozoic, but not enormously
so.

\section{Conclusions}
\label{concs}
Many authors have noted that, although the fossil record shows evidence of
events such as mass extinctions which are surely the result of unique and
specific causes, there also appear to be clear trends in the pattern of
origination and extinction.  Examples are the apparent decrease in the rate
of extinction and increase in the standing diversity over time.  Also
survivorship curves and the distribution of the lifetimes of taxa seem to
follow quite well-defined forms.  Several attempts have been made to
explain these results, many based on detailed assumptions about possible
evolutionary or ecological mechanisms underlying the processes of
extinction and origination.  In this paper we have taken the opposite
approach and explored instead the simplest possible theoretical framework.
We have shown that, by making only two basic assumptions about the fossil
extinction record, it is possible to place quite stringent limits on the
way in which extinction and origination can take place.  Our principal
results are as follows.
\begin{enumerate}
\item The fractional (or percentage) extinction rate at should decline over
  time at all taxonomic levels, but it cannot decline faster than $1/t$.
  This prediction is in good agreement with fossil record, which shows a
  decline close to this limiting form, within the statistical errors on the
  data.
\item The total extinction rate also can decline no faster than $1/t$, and
  empirically it too appears to be close to this limiting form.
\item The total origination rate must follow the extinction rate closely.
  For families, for example, the two can differ by no more than about
  1~family per~My.
\item Survivorship curves should fall off with taxon lifetime $\tau$ at
  least as fast as a power of $t/(t+\tau)$ and at most exponentially.  This
  result also is in good agreement with fossil data, and in particular with
  the empirical result known as Van Valen's law.
\item The diversity of taxa can increase at most linearly in time.  This is
  at odds with the conjecture of Benton~(1995) which states that diversity
  increases exponentially.  However, the data for diversity are
  sufficiently noise-ridden that a linear fit is perfectly acceptable.
\item The distribution of lifetimes of taxa must fall off as a power-law or
  faster.  This too appears to be consistent with fossil data, although the
  data in this case are rather poor.
\end{enumerate}

Our theory leaves many things unexplained.  Although we have been able to
place limits on the way in which various quantities may vary in time, we
have not been able to pin any of them down completely.  Furthermore, we
have offered no explanation of the assumptions of our theory, that
extinction declines on average, diversity increases, and no species lives
forever.  These issues cannot be resolved solely by mathematical analysis
of the kind presented here.  Only by exploiting real biological and
palaeontological insight can we hope to find a solution to these
long-standing problems.

\section{Acknowledgements}
The authors would like to thank Jack Sepkoski for providing the fossil data
for this study and Gunther Eble for useful discussions.  This work was
funded by the Santa Fe Institute and DARPA under grant number ONR
N00014--95--1--0975 and by a block grant from Statens Naturvidenskabelige
Forskingsr\aa d.

\def\refer#1#2#3#4#5#6{\item{\frenchspacing\sc#1}\hspace{4pt}
                       #2\hspace{8pt}#3  {#4} {\bf#5}, #6.}
\def\bookref#1#2#3#4{\item{\frenchspacing\sc#1}\hspace{4pt}
                     #2\hspace{8pt}{\it#3}  #4.}

\section*{References}
\baselineskip=15pt

\begin{list}{}{\leftmargin=2em \itemindent=-\leftmargin%
\itemsep=3pt \parsep=0pt \small}

\bookref{Bailey, N. T. J.}{1964}{The elements of stochastic
  processes.}{Wiley (New York)}

\refer{Baumiller, T. K.}{1993}{Survivorship analysis of Paleozoic
  Crinoidia.}{\it Paleobiology\/}{19}{304--321}

\bookref{Benton, M. J.}{1993}{The Fossil Record 2.}{Chapman and Hall
  (London)}

\refer{Benton, M. J.}{1995}{Diversification and extinction in the
  history of life.}{\it Science\/}{268}{52--58}
  
\refer{Flessa, K. W. \& Jablonski, D.}{1985}{Declining Phanerozoic
  background extinction rates: Effect of taxonomic structure?}{\it
  Nature\/}{313}{216--218}

\refer{Gilinsky, N. L.}{1994}{Volatility and the Phanerozoic decline of
  background extinction intensity.}{\it Paleobiology\/}{20}{445--458}

\item {\sc Newman, M. E. J. \& Eble, G. J.}\ \ 1998\ \ Decline in
  extinction rates and scale invariance in the fossil record.
  Submitted to {\it Paleobiology.}

\refer{Pease, C. M.}{1992}{On the declining extinction and origination rates
  of fossil taxa.}{\it Paleobiology\/}{18}{89--92}

\refer{Raup, D. M.}{1978}{Cohort analysis of generic survivorship.}{\it
  Paleobiology\/}{4}{1--15}

\item {\sc Raup, D. M.}\ \ 1988\ \ Testing the fossil record for
  evolutionary progress.  In {\it Evolutionary Progress,} M. H. Nitecki
  (ed.), University of Chicago Press (Chicago).

\refer{Raup, D. M.}{1991}{A kill curve for Phanerozoic marine species.}{\it
  Paleobiology\/}{17}{37--48}

\refer{Raup, D. M. \& Sepkoski, J. J., Jr.}{1982}{Mass extinctions in the
  marine fossil record.}{\it Science\/}{215}{1501--1503}

\refer{Raup, D. M. \& Sepkoski, J. J., Jr.}{1984}{Periodicity of
  extinctions in the geologic past.}{\it
  Proc. Natl. Acad. Sci.}{81}{801--805}

\item {\sc Sepkoski, J. J., Jr.}\ \ 1992\ \ A compendium of fossil marine
  animal families, 2nd edition.  {\it Milwaukee Public Museum Contributions
    in Biology and Geology\/} {\bf83}.

\refer{Sepkoski, J. J., Jr.}{1998}{Rates of speciation in the fossil
  record.}{\it Phil. Trans. R. Soc. London B\/}{353}{315--326}

\refer{Sibani, P., Schmidt, M. R. \& Alstr\o{}m, P.}{1995}{Fitness
  optimization and decay of extinction rate through biological
  evolution.}{\sl Phys. Rev. Lett.}{75}{2055--2058}

\refer{Sneppen, K., Bak, P., Flyvbjerg, H. \& Jensen,
  M. H.}{1995}{Evolution as a self-organized critical
  phenomenon.}{\it Proc. Natl. Acad. Sci.}{92}{5209--5213}

\refer{Valentine, J. W.}{1970}{How many marine invertebrate fossil
  species?}{\it J. Paleontology\/}{44}{410--415}

\refer{Van Valen, L. M.}{1973}{A new evolutionary law.}{\it Evol.
  Theory\/}{1}{1--30}

\refer{Van Valen, L. M. \& Maiorana, V.}{1985}{Patterns of
  origination.}{\it Evol. Theory\/}{7}{107--125}

\end{list}

\end{document}